\documentclass[onecolumn,showpacs,preprintnumbers,amsmath,amssymb,prc]{revtex4}

\usepackage{graphicx}
\usepackage{dcolumn}
\usepackage{bm}
\usepackage{color}
\begin{document}


\title{Elastic photonuclear cross sections for bremsstrahlung from relativistic ions}
\author{Rune E. Mikkelsen} \altaffiliation{Department of Physics and Astronomy, Aarhus University, Denmark.} \author{Allan H. S{\o}rensen} \altaffiliation{Department of Physics and Astronomy, Aarhus University, Denmark.}\author{Ulrik I. Uggerh{\o}j} \altaffiliation{Department of Physics and Astronomy, Aarhus University, Denmark.}
\date{\today}


\begin{abstract}
In this paper, we provide a procedure to calculate the bremsstrahlung spectrum for virtually any relativistic bare ion with charge 6$e$ or beyond, $Z\ge 6$, in ultraperipheral collisions with target nuclei. 
We apply the Weizs\"{a}cker-Williams method of virtual quanta to model the effect of the distribution of nuclear constituents on the interaction of the ion with the radiation target. 
This leads to a bremsstrahlung spectrum peaking at $2\gamma$ times the energy of the giant dipole resonance ($\gamma$ is the projectile energy in units of its rest energy).
A central ingredient in the calculation is the cross section for elastic scattering of photons on the ion. This is only available in the literature for a few selected nuclei and, usually, only in a rather restricted parameter range. Hence we develop a procedure applicable for all $Z\geq6$ to estimate the elastic scattering. The elastic cross section is obtained at low to moderate photon energies, somewhat beyond the giant dipole resonance, by means of the optical theorem, a dispersion relation, and data on the total absorption cross section.  The cross section is continued at higher energies by invoking depletion due to loss of coherence in the scattering. Our procedure is intended for any ion where absorption data is available and for moderate to high energies, $\gamma \gtrsim 10$. 
\end{abstract}

\maketitle

\section{Introduction}
When a charged particle penetrates a material, deflection in the electromagnetic field of target atoms causes emission of photons. We shall label this emission \textit{bremsstrahlung} provided the original particle is left intact. For light particles like the electron, bremsstrahlung is the dominant source of energy loss when traveling at sufficiently high relativistic energies. 
While the process is also relevant for other light particles like muons \cite{Groom2001}, it's less so for protons \cite{vanG1986}.
Bremsstrahlung from heavy charged particles penetrating matter at high energies has never been measured. 
If the projectile is treated as a point charge, classical and quantal calculations indicate that bremsstrahlung should be a major energy-loss channel for highly relativistic bare ions and, as for electrons, ultimately the dominant one. However, when due account is taken for the internal structure of the nucleus and the latter is required to remain intact, which restricts to ultraperipheral collisions with target nuclei, the emission ends up confined to relatively soft quanta with the result that bremsstrahlung never adds substantially to the energy loss \cite{Sore10}, \cite{JenSor2013}.

We have previously determined the bremsstrahlung spectrum for bare lead ions penetrating a lead target at relativistic energies (typically $\gamma$ =170 and beyond). Our calculation was based on the Weizs\"{a}cker-Williams method of virtual quanta. There are four contributions, the main contribution derives from scattering of virtual photons of the target nucleus on the incoming ion. The scattering is determined in the rest frame of the projectile and the main bremsstrahlung component is obtained by subsequent transformation to the target frame. Obviously, a central ingredient is the photo-nuclear scattering cross section. For the case of lead experimental data for elastic photon scattering are available; these data were used to construct a relatively accurate fit which served as input in the calculation of bremsstrahlung for a lead ion \cite{Sore10}. 
The nature of the photon-nucleus interaction is reflected in the structure of the bremsstrahlung cross section, which shows a significant peak at, roughly, 25$\gamma$ MeV, enhanced above the results for a pointlike source of the same charge and mass. 

Computation of the bremsstrahlung spectrum for heavy ions other than lead is complicated by the fact that data for elastic photon scattering is generally unavailable. The elastic cross section therefore has to be estimated by other means. For such cases, we model the elastic cross section from the total absorption cross section by application of the optical theorem and a dispersion relation at energies up to and somewhat beyond the giant dipole resonance and include depletion at higher energies based on the underlying physics. We adjust so as to obtain a single closed procedure applicable for all nuclei by requiring reproduction at an acceptable level of elastic scattering data available for C, O, and Pb.

While the theoretical results for lead ions have awaited experimental data to check against, other relevant experiments may soon be underway. 
Starting in 2015, Ar, Xe and Pb beams will be available at the Super Proton Synchrotron at CERN for fixed-target experiments. We shall therefore apply our procedure to the case of argon.\\

As discussed below, the relative widths of the expected radiation peaks give indirect information on the Giant Dipole Resonance of the projectile. These widths thus inform about the symmetries of the nucleus - if it is spherical or not, for instance. The level of accuracy with which this information can be obtained by such an indirect method is not competitive in the case of stable elements, but in the case of bremsstrahlung emission it may yield information even for nuclei with lifetimes down to the picosecond regime.

The reader should note that the nuclear community applies the term bremsstrahlung much broader than we do here. It is used for photon emission even in energetic central collisions where both collision partners disintegrate. The more restrictive use of the term here follows from our perspective, viz. the slowing down of charged particles.
\section{Bremsstrahlung in ultraperipheral collisions}
\label{sec:BS}
The classical radiation cross section for a pointlike particle serves as a reference for our bremsstrahlung results. For a pointlike bare relativistic ion with charge $Ze$, nucleon number $A$, and mass $M = A M_u$ penetrating a target of atomic number $Z_t$ at energy $E$, this cross section is given as \cite{Jackson75} 
\begin{equation}
\frac{d \chi}{d \hbar\omega} = \frac{16}{3}\frac{Z_t^2Z^4}{A^2}\alpha r_u ^2L \; ,
\label{eq:Reference}
\end{equation}
where $\alpha \equiv e^2/\hbar c^2$ is the fine-structure constant, $r_u \equiv e^2/M_uc^2$ the classical nucleon radius, and $M_u$ the atomic mass unit. Note that $\chi$ carries the dimension of energy times area; a count spectrum is obtained by dividing by the photon energy $\hbar\omega$.
The factor $L$ appearing in Eq. (\ref{eq:Reference}) is given by
\begin{align}
\begin{split}
 L  \approx \textrm{ln} \left( \frac{233M}{Z_t^{1/3}m} \right) - \frac{1}{2} \left[ \textrm{ln} \left(1+r^2 \right) - \frac{1}{1+r^{-2}} \right], \\
 r = \frac{96 \hbar \omega}{\gamma \gamma_- Z_t^{1/3}mc^2},
\end{split}
\end{align}
where $m$ denotes the electron mass, $\gamma \equiv E/Mc^2$, and $\gamma_- \equiv (E- \hbar\omega)/Mc^2$. 
It is essentially the logarithm of the ratio of the effective maximum and minimum momentum transfers to the scattering center. 
The logarithmic factor $L$ represents the only energy-dependent part of the radiation cross section (\ref{eq:Reference}) which thereby is almost constant up to the primary energy $E$ (neglecting quantum recoil).

To obtain the bremsstrahlung emitted by a composite nucleus penetrating a target at relativistic energy, the Weizs\"{a}cker-Williams method of virtual quanta was applied in \cite{Sore10}. The nuclear structure plays an essential role in the photon emission and leads to a significant energy-dependence of the cross section. 
In the Weizs\"{a}cker-Williams approach, the bremsstrahlung stems from a combination of four sources: scattering of virtual photons of the target nucleus on the projectile, scattering of virtual photons of the target electrons on the projectile, scattering of virtual photons of the projectile on the target nucleus, and scattering of the virtual photons of the projectile on target electrons \cite{JenSor2013}. 
The main component is due to scattering of virtual photons from target nuclei on the projectile. 
When these scatter at large angles they undergo a Lorentz boost that increases their energy by $\approx 2\gamma$ easily resulting in the emission at GeV energies for virtual photons incident at 10--20 MeV on relativistic projectiles with $\gamma$ of the order of $10^2$.
 
For the main bremsstrahlung component, the scattering of the virtual photons of the target nucleus on the projectile is considered in the rest frame of the latter, where variables are denoted by primes.
By multiplying the elastic photon scattering cross section differential in scattering angle, $d\sigma /d\Omega'$, with the spectrum of virtual photons differential in photon energy, $dI'/d\hbar\omega'$, we obtain the doubly-differential radiation cross section
\begin{equation}
\frac{d^2\chi '}{d\hbar\omega'd\Omega'} = \frac{d\sigma}{d\Omega'}\frac{dI'}{d\hbar\omega'} \; .
\label{eq:doubleDiff}
\end{equation}
Explicit expressions for the Weizs\"{a}cker-Williams photon intensity, $dI'/d\hbar\omega'$, are given in Refs. \cite{Sore10}, \cite{JenSor2013}. We shall apply the intensity pertaining to an exponentially screened Coulomb potential of the target nuclei. Determination of the elastic cross section $d\sigma /d\Omega'$ is the major issue of this paper; it is discussed in Section \ref{sec:Elastic} below.\\

To obtain a measurable bremsstrahlung yield, the radiation cross section (\ref{eq:doubleDiff}) has to be transformed to the laboratory frame (rest frame of the target). For this, use is made of the invariance relation \cite{Jackson75,Sore10}
\begin{equation}
\frac{1}{\omega^2}\frac{d^2\chi}{d\hbar\omega d\Omega} = \frac{1}{\omega'^2}\frac{d^2\chi '}{d\hbar\omega'd\Omega'} \; .
\label{eq:transform}
\end{equation}
We obtain the bremsstrahlung spectrum in the laboratory, differential in energy and solid angle, by multiplication of Eq. (\ref{eq:transform}) by $\omega^2$ and by expressing the photon energy and scattering angle in the projectile rest frame, appearing on the right-hand-side of the equation, in terms of their laboratory equivalents. From the relativistic Doppler (or Lorentz) transformation we have
\begin{equation}
\omega' = \gamma\omega\left(1-\beta\cos\theta\right)
\label{eq:DopplerE}
\end{equation}
and
\begin{equation}
\cos\theta' = \frac{\cos\theta - \beta}{1- \beta\cos\theta}\; ,
\label{eq:angle}
\end{equation} 
where $\theta$ is the angle between the direction of photon emission and the direction of the projectile and $\beta = v/c$ is the projectile speed relative to that of light. 

\begin{figure}[hbt]
\begin{center}
\includegraphics[width=0.5\textwidth]{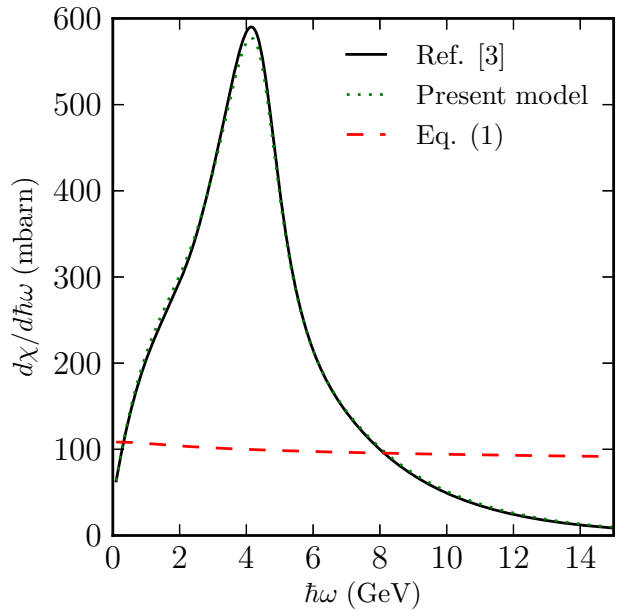}
\caption{\textsl{Bremsstrahlung spectrum for a bare lead nucleus penetrating a lead target at a $\gamma$-value of 170. The solid black line is adapted from \cite{Sore10}, the green dotted line shows the result of the present approach, Sec. \ref{subsec:newapproach}, while the red dashed line shows the reference cross section (\ref{eq:Reference}).}
\label{fig:BSPb170}}
\end{center}
\end{figure}

Equations (\ref{eq:doubleDiff}--\ref{eq:angle}) summarize the necessary ingredients to calculate the bremsstrahlung spectrum; additional information as well as useful and quite accurate approximations relying on high values of $\gamma$ may be found in Ref. \cite{Sore10}. 
Figure \ref{fig:BSPb170} displays the result for the main bremsstrahlung component, as a function of photon energy, for a bare lead ion penetrating a lead target at $\gamma =170$. Among the significant features are the peak near 4 GeV and the subsequent fall-off. None of these appear for a pointlike object which produces an essentially flat and featureless spectrum. The peak and the following depletion for $^{208}$Pb reflects the structure of the elastic scattering cross section presented below, the peak appearing near $2\gamma$ times that of the giant dipole resonance in the cross section.\\

\section{Elastic photonuclear cross sections}
\label{sec:Elastic}
To calculate bremsstrahlung according to the Weizs\"{a}cker-Williams scheme we need the elastic photonuclear cross section. 
This is only available in the literature for a limited number of nuclei, and even then perhaps only in the form of a few data points for a few scattering angles if not just one. To compute the bremsstrahlung spectrum for a wider range of ions, including Ar, we hence seek a general procedure to determine the elastic photonuclear cross section.

\subsection{Previously applied procedures}

\subsubsection{Lead}
\label{subsubsec:LeadElastic}

\begin{figure*}[hbt]
\begin{center}
\includegraphics[width=\textwidth]{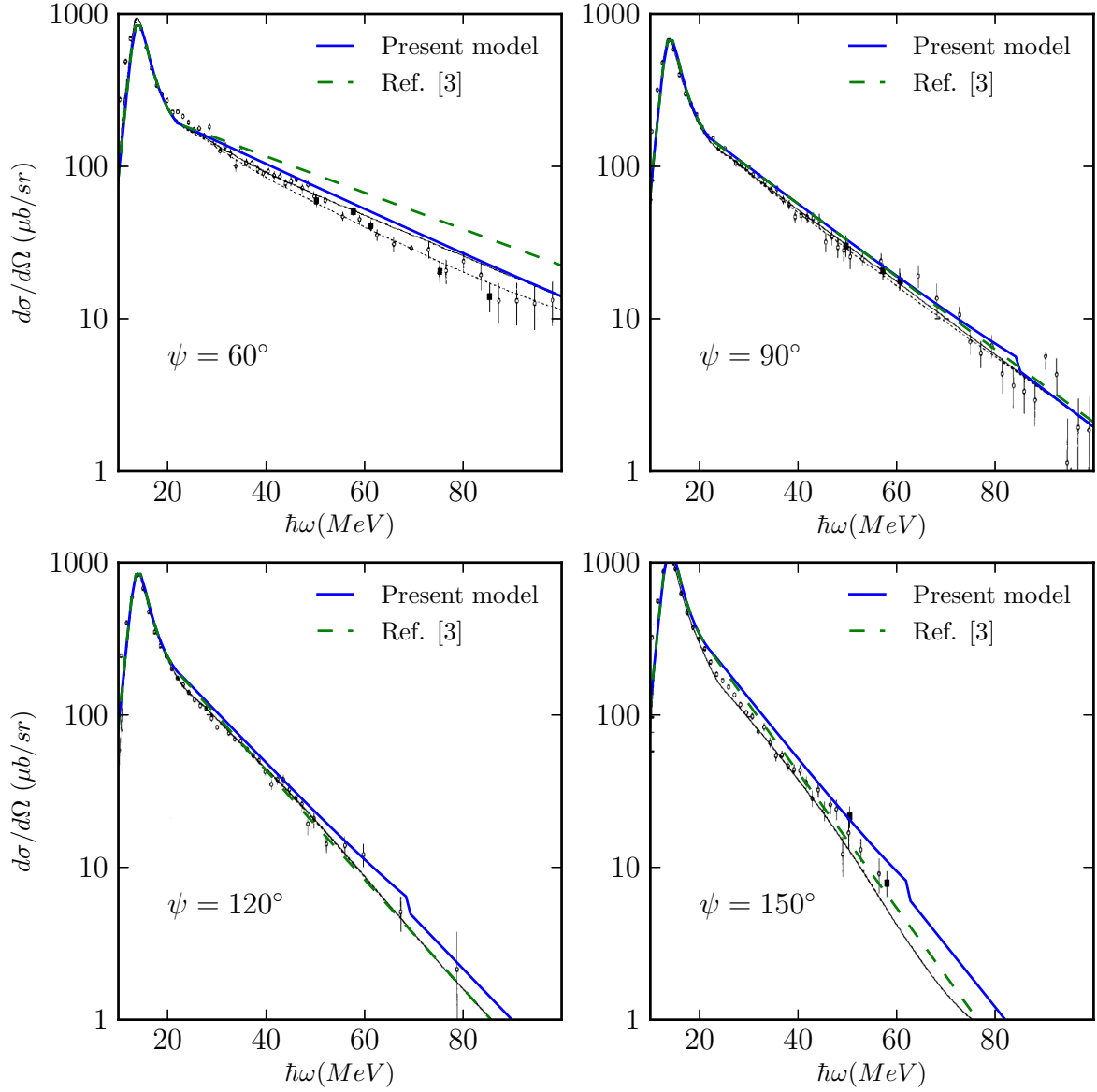}
\caption{\textsl{Elastic photonuclear cross section for lead for four different scattering angles. The variation with photon energy well beyond the giant dipole resonance is included. Data and curves drawn in black are adapted from Schelhaas \textit{et al.}\cite{Schelhaas88}; see text for further explanation. The green dashed curve displays the fit (\ref{eq:Pbfit}) whereas the solid curve drawn in blue shows the result of the present approach, Sec. \ref{subsec:newapproach}. }} 
\label{fig:oldPbfit}
\end{center}
\end{figure*}

Lead is a special case since extensive data for elastic photon scattering exist  \cite{Schelhaas88}.
In \cite{Sore10} a fit to these data was constructed for the purpose of calculating the bremsstrahlung spectrum. This fit, applied to obtain the solid black curve in Fig. \ref{fig:BSPb170}, assumes the form \cite{Sore10}
\begin{equation}
   \frac{d\sigma}{d\Omega'} = Z^2r_p^2\frac{1}{2} \left( 1+ \cos^2\psi'\right) \times \\ 
   \begin{cases}
   \left(\frac{ZM_p}{M} \right)^2; & \hbar\omega'  < \hbar \omega_1 \\
          0.793 \frac{\left( \hbar\omega'  \right)^4}{\left( \left( \hbar\omega' \right)^2-\left( E_m \right)^2\right)^2 + \left( \Gamma\hbar\omega' \right)^2}; & \hbar\omega_1 < \hbar\omega'  < \hbar\tilde{\omega}_2\\
          1.93 \exp \left(-\epsilon \left( \hbar\omega'-\hbar\tilde{\omega}_2 \right) \sin^2 \frac{\psi'}{2}\right); & \hbar\tilde{\omega}_2  < \hbar \omega' 
  \end{cases}
  \label{eq:Pbfit}
\end{equation}
where the scattering angle $\psi'$ of the photon relates to the emission angle $\theta'$ relative to the direction of projectile motion as $\psi' = \pi - \theta'$. We keep primes on all variables since, eventually, the differential cross section will be applied in the rest frame of the projectile. 
The quantity $r_p\equiv e^2/M_pc^2$ is the classical radius of the proton (mass $M_p$) and $E_m=13.7$ MeV, $\Gamma =4.19$ MeV. 
Clearly, for lead the atomic number $Z$ assumes the value 82, but the symbol is kept to underline the physical ingredients of the components:
The upper line in (\ref{eq:Pbfit}) corresponds to scattering on a single rigid object of charge $Ze$ and mass $M$, the middle line reflects resonant coherent scattering on $Z$ quasi-free protons, the lower line effectively imposes restriction to scattering angles sufficiently small that coherent interaction with all protons is ensured. The lower dividing energy $\hbar \omega_1$ essentially reflects typical excitation or binding energies; a value of 7.69 MeV ensures continuity of the fit. The upper dividing energy corresponds roughly to $\hbar \omega_2\equiv\hbar c/R=28.2$ MeV where the (reduced) wavelength of the photon equals the nuclear radius $R$; a value of $\hbar\tilde{\omega}_2=22.0$ MeV was found to provide a good fit. Details may be found in \cite{Sore10}. Fig. \ref{fig:oldPbfit} displays the quality of the fit. Despite its simplicity the fit reproduces the variation of the data with energy and angle quite well, with a tendency to overshoot at small angles for high energies.

\subsubsection{Modified dispersion relation}
\label{subsubsec:modDispRel}
The differential cross section for forward elastic scattering can be calculated from the total photo cross section by application of the optical theorem and a dispersion relation. 
This is beneficial since the total cross section essentially is the same as the absorption cross section over a wide range of energies, absorption is comparatively easier to measure than elastic scattering, and theoretically calculated values of the total absorption cross section are readily available for a variety of projectiles for particle penetration purposes. To obtain the full cross section for elastic scattering requires additional input. Below, we shall outline some approaches taken by other authors.
All variables in this subsection refer to the rest frame of scatterer and for convenience we shall deviate from our standard notation and omit primes on this singular occasion.

The optical theorem rests on conservation of probability. The theorem relates the total cross section for photon interaction $\sigma_{\text{tot}}$ at given photon energy $\hbar\omega$ to the imaginary part of the forward scattering amplitude, R$\left( \hbar \omega ,\psi=0 \right)$, where $\psi$ signifies the scattering angle  \cite{Sakurai}. Over most of the energy interval relevant below, $\sigma_{\text{tot}}$ is very nearly equal to the total absorption cross section $\sigma_{\text{a}}$ which hence will be substituted in its place. With this approximation the optical theorem reads
\begin{equation}
\textrm{Im}R(\hbar \omega,0)=\frac{\omega}{4\pi c}\sigma_{\text{a}} \; .
\label{eq:Im}
\end{equation}
Causality and analyticity leads, for the case where the scattering amplitude does not vanish in the limit $\hbar\omega\rightarrow 0$, to the dispersion relation
\begin{equation}
\mathrm{Re} R(\hbar \omega ,0) = R(0,0)+\frac{2\omega^2}{\pi} 
P \int_0^{\infty} \frac{d\omega' \,\mathrm{Im}R(\hbar \omega',0)}{\omega'(\omega'^2-\omega^2)} \; ,
\label{eq:ReGeneral}
\end{equation}
where $P$ denotes the principal value. By application of Eq. (\ref{eq:Im}) and substitution of the Thomson amplitude
\begin{equation}
    D=-\frac{(Ze)^2}{Mc^2}=-\frac{Z^2e^2}{AM_uc^2}=-\frac{Z^2}{A}r_u
\label{eq:D}
\end{equation}
for $R(0,0)$ the dispersion relation transforms to
\begin{equation}
\mathrm{Re}R(\hbar \omega ,0) = D + \frac{\omega^2}{2\pi^2c} 
P \int \frac{d\omega' \sigma_{\text{a}}(\hbar \omega')}{\omega'^2-\omega^2}\; .
\label{eq:Re}
\end{equation}
The forward scattering amplitude, and hence the differential cross section for forward scattering, may be obtained by application of Eqs. (\ref{eq:Im}) and (\ref{eq:Re}).

The relations above give no information on the elastic scattering cross section at finite angles. But if for other reasons a particular energy-independent angular variation $g(\psi)$ may be assumed, we can compute the full differential cross section as
\begin{equation}
\frac{d\sigma}{d\Omega}(\hbar \omega ,\psi) = \lvert R\left(\hbar \omega,0\right) \rvert^2 g(\psi) \; .
\label{eq:ImAndRe}
\end{equation}
For a dipole radiation pattern the angular function assumes the familiar form 
\begin{equation}
g(\psi) = (1 + \cos^2\psi)/2 \; , 
\label{eq:gdipole}
\end{equation}
cf. Eq. (\ref{eq:Pbfit}).

Application of Eqs. (\ref{eq:Im}), (\ref{eq:D})--(\ref{eq:gdipole}) results in elastic scattering cross sections in excess of the available experimental data at energies above the giant dipole resonance. 
The calculated values overestimate the scattering cross section beyond the resonance because the inherent assumption of coherent action of all constituents of the nucleus does not hold.  
Various authors have invoked modifications of the procedure described above to better account for the experimental data. 
One approach has been to change the Thomson amplitude (\ref{eq:D}) through the introduction of one or more form factors.
Ziegler \cite{Ziegler} for example, in a study of the elastic scattering of photons off lead at energies around and above the giant dipole resonance, essentially matched experimental data by modifying the Thomson amplitude to 
\begin{equation}
D'(\hbar \omega,\psi) = 
   \left[ \left(1 + \kappa_{\text{res}} \right) \frac{NZ}{A} - ZF_Z(q) - \kappa_{\text{res}}\frac{NZ}{A}F_{\text{ex}}(q) \right] r_u \; .
\label{eq:Dmod}
\end{equation}
The reader is referred to \cite{Ziegler} for definition of the quantities entering on the right-hand-side of Eq. (\ref{eq:Dmod}). Here it suffices to note that the two form factors $F_Z$ and $F_{\text{ex}}$ depend on the momentum transfer $\hbar q$,
whereby substitution of $D'$ (\ref{eq:Dmod}) for $D$ (\ref{eq:D}) in the dispersion relation (\ref{eq:Re}) introduces an angular dependence in the cross section (\ref{eq:ImAndRe}) beyond that dictated by $g(\psi )$. 
This shows at energies typically beyond the giant dipole resonance as a reduction which brings the theoretical values in proximity of the experimental data. 
In Figure \ref{fig:oldPbfit}, reprinted from \cite{Schelhaas88}, we show data and calculated values for the cross sections of elastic scattering of photons off $^{208}$Pb for four different angles.
The data as well as the curves drawn in black appeared in the original publication \cite{Schelhaas88}.  
The full drawn black curves show how using $D'$ allows a good reproduction of the experimental data. The dotted black curves displays the result of the calculation when the mesonic component of $D'$ has been reduced by 70 $\%$. Since changes are limited in the considered range of parameters, so is the importance of the mesonic channel.
While the reproduction of experimental data is quite successful it should be noted that the procedure is inconsistent in so far as an angular dependence is introduced in the dispersion relation despite the fact that the latter solely concerns the singular case $\psi=0$.

The procedure just outlined has only been applied to calculate the elastic photon scattering cross sections for lead.
For lower charge ions like $^{12}$C and $^{16}$O, other authors, \cite{Hayward83} and \cite{Hayward92}, have studied alternative modification schemes.
These include modification of both the Thomson amplitude and the dispersion relation. 
In Figures \ref{fig:compareCarbon} and \ref{fig:compareCarbonHighE}, reprinted from \cite{Hayward83} and \cite{Hayward92}, we display data along with theoretical values for $^{12}$C (data and solid black curves appeared in the original publications) and in Figure \ref{fig:Oxygen} for $^{16}$O from \cite{Hayward83}. 
As for the case of lead the very idea of introducing finite angles in the dispersion relation for the forward scattering amplitude combined with the optical theorem is inconsistent and will thus be abandoned below. Since the optical theorem and the dispersion relation are rooted in very fundamental physical principles, conservation of probability and causality, we shall maintain both of them without modification and introduce depletion at finite angles for energies beyond the giant dipole resonance by other means. 
 
\subsection{New approach}
\label{subsec:newapproach}
The use of the optical theorem, the dispersion relation, and the dipole radiation pattern produces results in quite reasonable agreement with experimental data for elastic photon scattering for energies around the giant dipole resonance and somewhat beyond. 
Hence, for energies below, up to, and slightly above the giant dipole resonance we use Eqs. (\ref{eq:Im}--\ref{eq:gdipole}) strictly as they appear to get the elastic cross section. 
To obtain the necessary depletion at higher energies we employ three central guidelines: we wish to keep it simple, it must be internally consistent and it has to be motivated by physics. That is, we introduce modifications based on clear physical principles and modelled fairly roughly, yet without deteriorating the possibility of obtaining reliable bremsstrahlung spectra.
In the construction, we shall draw on the fact that we have a procedure which works well for $^{208}$Pb. 
\begin{figure*}[hbt]
\begin{center}
\includegraphics[width=\textwidth]{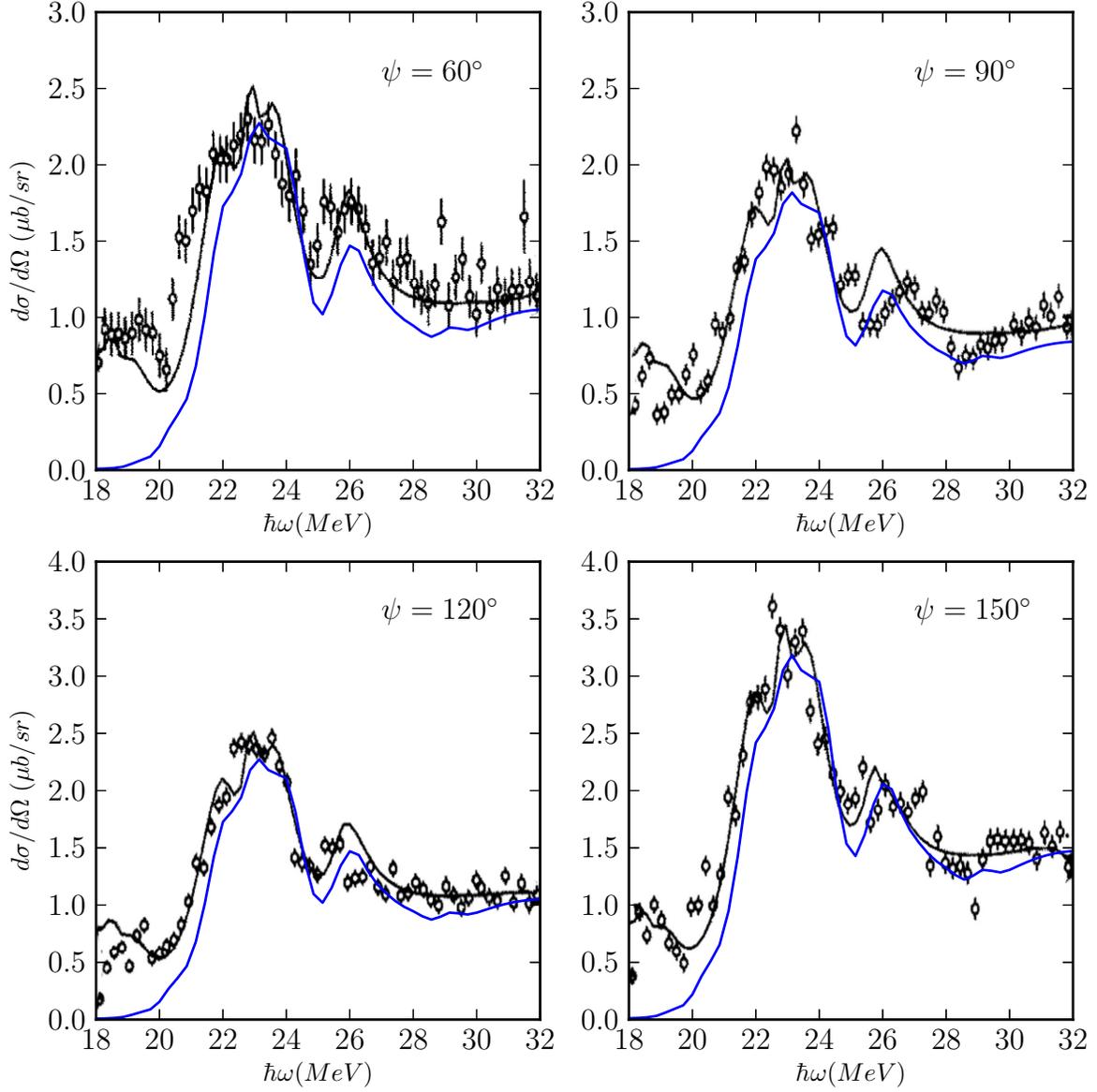}
\caption{\textsl{Comparison of our calculation with data on the elastic-plus-inelastic cross sections for $^{12}C$ from \cite{Schelhaas90} for four different scattering angles. Data points and the dark line are reprinted from the original paper. Our calculations somewhat undershoot the data, which is expected since we only consider the elastic part.}}
\label{fig:compareCarbon}
\end{center}
\end{figure*}

Fall-off of the elastic cross section at high energies appears since, here, only a fraction of the protons act in coherence to scatter a photon of short wavelength. Depletion appears when the momentum transfer $\hbar q$ to the nucleus becomes larger than $\hbar /R$. Since 
\begin{equation}
 qR = 2\, \frac{\omega'}{\omega_2}\sin\frac{\psi'}{2} \; ,
\label{eq:qR}
\end{equation}
we search among depletion factors depending on photon energy $\hbar\omega'$ as well as on $\sin\frac{\psi'}{2}$. For lead we found the construct appearing in the lower line of (\ref{eq:Pbfit}) to work very well, except for slight deviations at moderate to small scattering angles. Hence, as a starting point we adapt the same form for other nuclei, that is,
\begin{equation}
\frac{d\sigma}{d\Omega'}\left( \psi' \right) = Z^2r_p^2\frac{1}{2} \left( 1+ \cos^2\psi'\right) \times 
F_0 \; , \hspace*{1.cm} \hbar\omega' > \hbar\tilde{\omega}_2 \; ,
\label{eq:depletion0}
\end{equation}
where
\begin{equation}
 F_0\equiv k\exp \left[-\epsilon(\hbar\omega'-\hbar\tilde{\omega}_2)\sin^2(\psi'/2)\right] \; .
\label{eq:F0}
\end{equation}
In doing so, the parameters $\epsilon$ and $\hbar\tilde{\omega}_2$ have to be adjusted. Application of Eqs. (\ref{eq:depletion0}--\ref{eq:F0}) reveals that the value of the former parameter should decrease with decreasing mass number $A$ while the latter should increase in order to reproduce the data. The factor $k$ appearing in the definition of $F_0$ is chosen such as to ensure continuity at the dividing energy,
\begin{equation}
k =\lvert R(\hbar\tilde{\omega}_2,0)\rvert^2/(Zr_p)^2 \; .
\label{eq:kcontinuity}
\end{equation}
In Table \ref{tab:k}, we provide the list of $k$-values that have been extracted for the different materials in this study. We note that the variation with $A$ is modest with $k$ assuming the value 2 within, roughly, 10 \% for the considered wide range of ions. We take this finding as support for the sanity of our rather rough model.

\begin{table}[htdp]
\caption{Values of parameter $k$ as defined in (\ref{eq:kcontinuity}) for different materials, with $\hbar\tilde{\omega}_2$ defined in (\ref{eq:valuedividingE}). The absorption data was taken from Ref. \cite{ENDF}.}
\begin{center}
\begin{tabular}{c|rrrr}
   & C & O & Ar & Pb       \\ 
\hline  
$k$ 	& $1.99$ & $1.82$ & $2.15$ & $2.27$
\end{tabular}
\end{center}
\label{tab:k}
\end{table}%

To get the depletion factor $F_0$ we argued on coherent scattering on all protons \cite{Sore10}, the underlying assumption being that if a photon scatters on a single proton, this proton recoils sufficiently violently to leave the nucleus. 
That is, the projectile breaks up and the event is not classified as bremsstrahlung. 
This is an exaggeration. The nucleus only breaks up if the photon transmits an energy to the proton in excess of its binding energy. Since typical binding energies are of order 8 MeV and the energy transfer assumes the value $(\hbar q)^2/2M_p$, the photon energy has to exceed, roughly, $60$ MeV$/\sin\frac{\psi'}{2}$ in order for the proton to leave the nucleus. In other words, for lower photon energies, also incoherent scattering on a single proton contributes to elastic scattering of the photon off the nucleus. 
The cross section for coherent scattering off $Z$ protons scales with $Z^2$, cf. (\ref{eq:Pbfit}), (\ref{eq:depletion0}) while the cross section for incoherent scattering scales as $Z\times 1^2=Z$. 
That is, since the relative weight of the incoherent contribution is $1/Z$, this correction is minute for Pb. 
However, for light targets as C an O it may well add 10--20 {\%}. We include the incoherent contribution approximately by replacing the depletion factor $F_0$ in Eq. (\ref{eq:depletion0}) by
\begin{equation}
 F_1\equiv F_0 + \frac{1}{Z}\left( 1-\frac{F_0}{k}\right) \Xi \; .
\label{eq:F1}
\end{equation}
Here the last factor is supposed to switch off the incoherent contribution when relevant. We take it simply as a step function, $\Xi =\Theta (60 \textrm{ MeV} - \hbar\omega'\sin\frac{\psi'}{2})$, where $\Theta(x)$ is the Heaviside function (1 for positive arguments, 0 otherwise).

\begin{figure}[hbt]
\begin{center}
\includegraphics[width=0.5\textwidth]{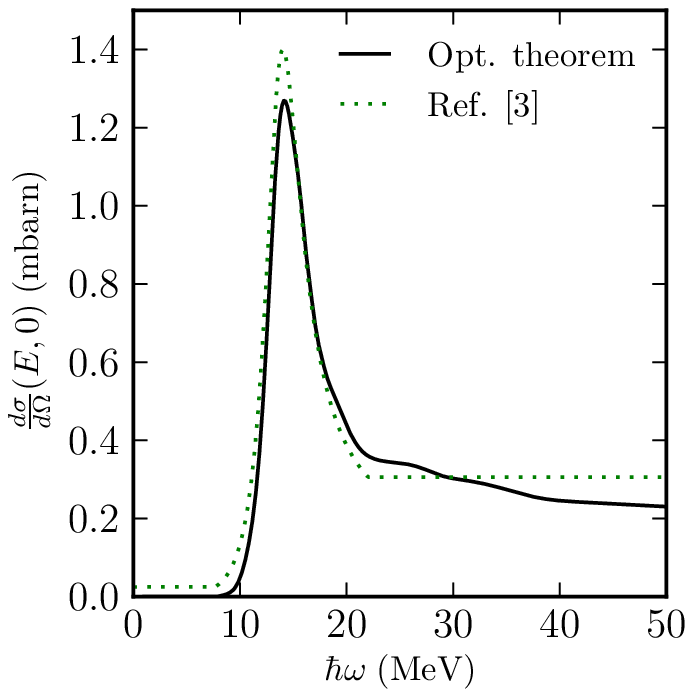}
\caption{\textsl{Forward scattering cross section for $^{208}$Pb. The solid dark line shows the result of applying the optical theorem and the dispersion relation using data from Ref. \cite{ENDF}. The dotted green line shows the fit from \cite{Sore10}. 
The dividing energy, $\hbar\tilde{\omega}_2$, for lead is $22$ MeV. 
}}
\label{fig:BaggSorDisp}
\end{center}
\end{figure}

If we choose the value of the damping parameter $\epsilon$ appearing in $F_0$, Eq. (\ref{eq:F0}), such as to reproduce the cross section at large scattering angles for one of the lighter nuclei, we end up with substantial errors at small angles. That is, the deviations which were moderate for Pb are magnified for C and O. The origin of the small-angle deviations is fairly obvious if we consider the cross section for forward scattering: while (\ref{eq:Pbfit}) predicts a constant beyond $\hbar\tilde{\omega}_2$, application of the optical theorem and the dispersion relation yields a cross section that falls off with increasing energy (as long as we stay well below the pion-production threshold of 140 MeV), cf. Fig. \ref{fig:BaggSorDisp}. This suggests including a term independent of scattering angle in the high-energy depletion factor. 
Hence we modify our previous expression (\ref{eq:F1}) to
\begin{equation}
 F\equiv F_0\exp \left[-\delta (\hbar\omega'-\hbar\tilde{\omega}_2)\right] + \frac{1}{Z}\left( 1-\frac{F_0}{k}\right)\Theta (60 \textrm{ MeV} - \hbar\omega'\sin\frac{\psi'}{2})  \; .
\label{eq:Ffinal}
\end{equation}
This is our final expression which we shall use in place of $F_0$ in Eq. (\ref{eq:depletion0}) in order to get the elastic scattering cross section for photon energies beyond $\hbar\tilde{\omega}_2$.
The abrupt cut of the small incoherent contribution results in an artificial step in the elastic cross section which, however, is small and does not show in the bremsstrahlung cross section when integrated over emission angles.

\begin{figure*}[hbt]
\begin{center}
\includegraphics[width=\textwidth]{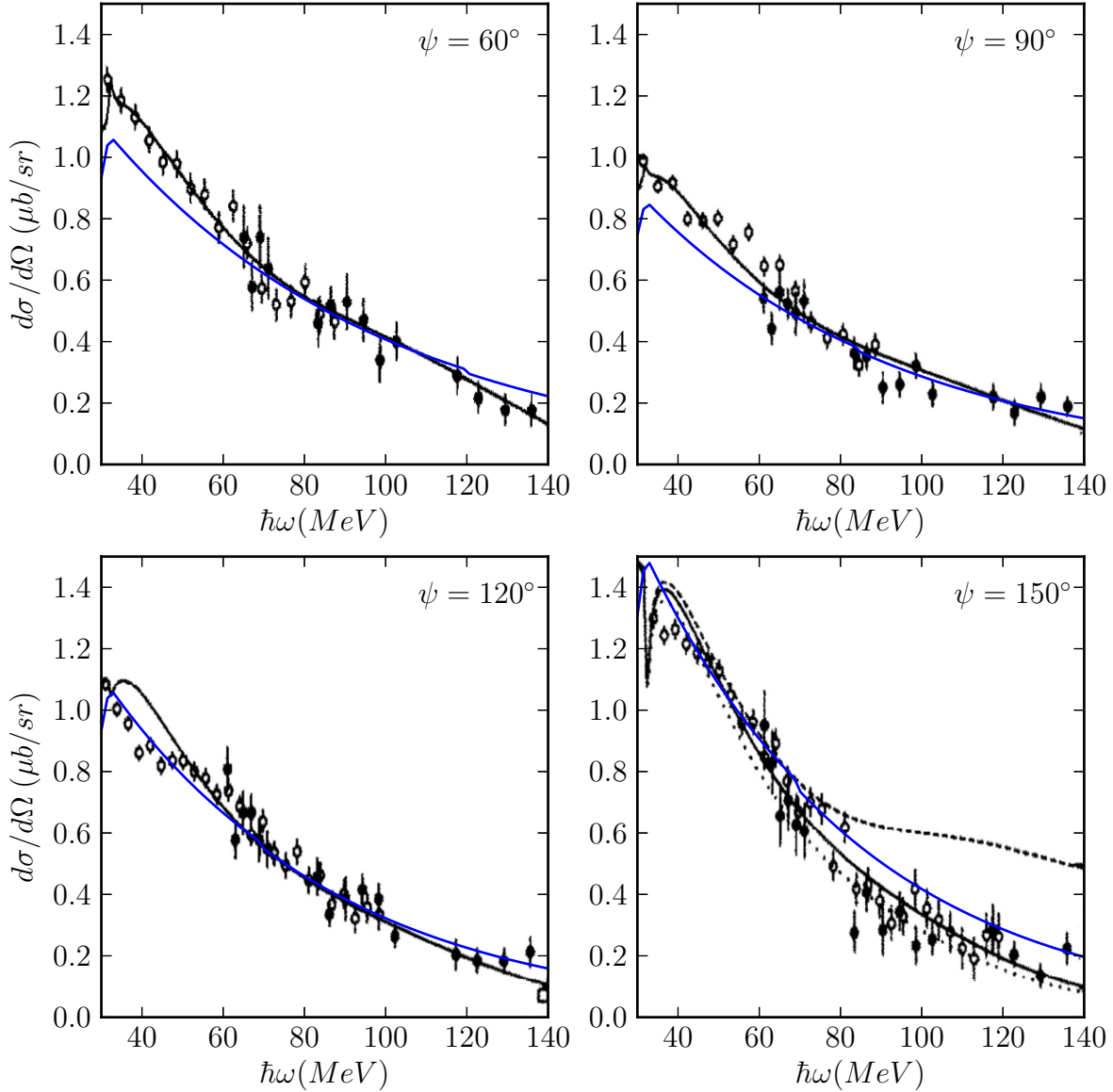}
\caption{\textsl{Elastic scattering cross sections for photons on $^{12}C$ for four different scattering angles. The blue line represents the present model and the dark lines and data points are reprinted from \cite{Schelhaas90}. Their data include elastic and inelastic contributions, whereas our model only describes the elastic scattering.}}
\label{fig:compareCarbonHighE}
\end{center}
\end{figure*}

The parameters 
appearing in Eqs. (\ref{eq:F0},\ref{eq:Ffinal}) generally depend on the mass number $A$. A comparison of the experimental data available for elastic scattering of photons on nuclei of C, O, and Pb reveals that an efficient fit may be obtained by choosing the energy $\hbar\tilde{\omega}_2$, which divides between the two theoretical constructs, to be off-set by a constant from the position of the giant dipole resonance.
The location of the latter can be parameterized as a function of 
$A$ \cite{Gaardhoeje92}.  Accordingly, our choice for the dividing energy is
\begin{equation}
\hbar\tilde{\omega}_2 [\textrm{MeV}] = 17 A^{-1/3} + 25 A^{-1/6} + 8.9 \; ,
\label{eq:valuedividingE}
\end{equation}
where the offset is chosen to be $8.9$ to arrive at $22.0$ MeV for lead.
In order to obtain values for $\epsilon$ and $\delta$ we estimate an exponential high-energy depletion coefficient $\alpha$ from the experimental data for a given target at given angle, that is, an energy dependence $\exp (-\alpha (\hbar\omega' - \hbar\tilde{\omega}_2))$ is assumed with $\alpha$ depending on target material and scattering angle. Values of $\epsilon$ and $\delta$ for the given target are extracted by assuming a linear dependence of $\alpha$ on $\sin^2(\psi'/2)$. Finally, values obtained for different targets are compared. For this analysis, which neglects the small incoherent contribution, we have only found sufficient experimental data for Pb and C. Hence the following "universal" values are, strictly speaking, only known to be justified for $Z=6$ and $Z=82$. Our results are:
\begin{equation}
\epsilon = 0.086 \times Z/82 
\label{eq:valueepsilon}
\end{equation}
and
\begin{equation}
\delta = 0.013 \; .
\label{eq:valuedelta}
\end{equation}
A dependence of $\epsilon$ on $Z$ rather than on $A$ is plausible since only the protons are involved in the scattering of the photon, be it coherent or incoherent. The target-independent value of $\delta$ hints at similarity of the forward scattering amplitude somewhat beyond the giant dipole resonance. 

\begin{figure*}[hbt]
\begin{center}
\includegraphics[width=\textwidth]{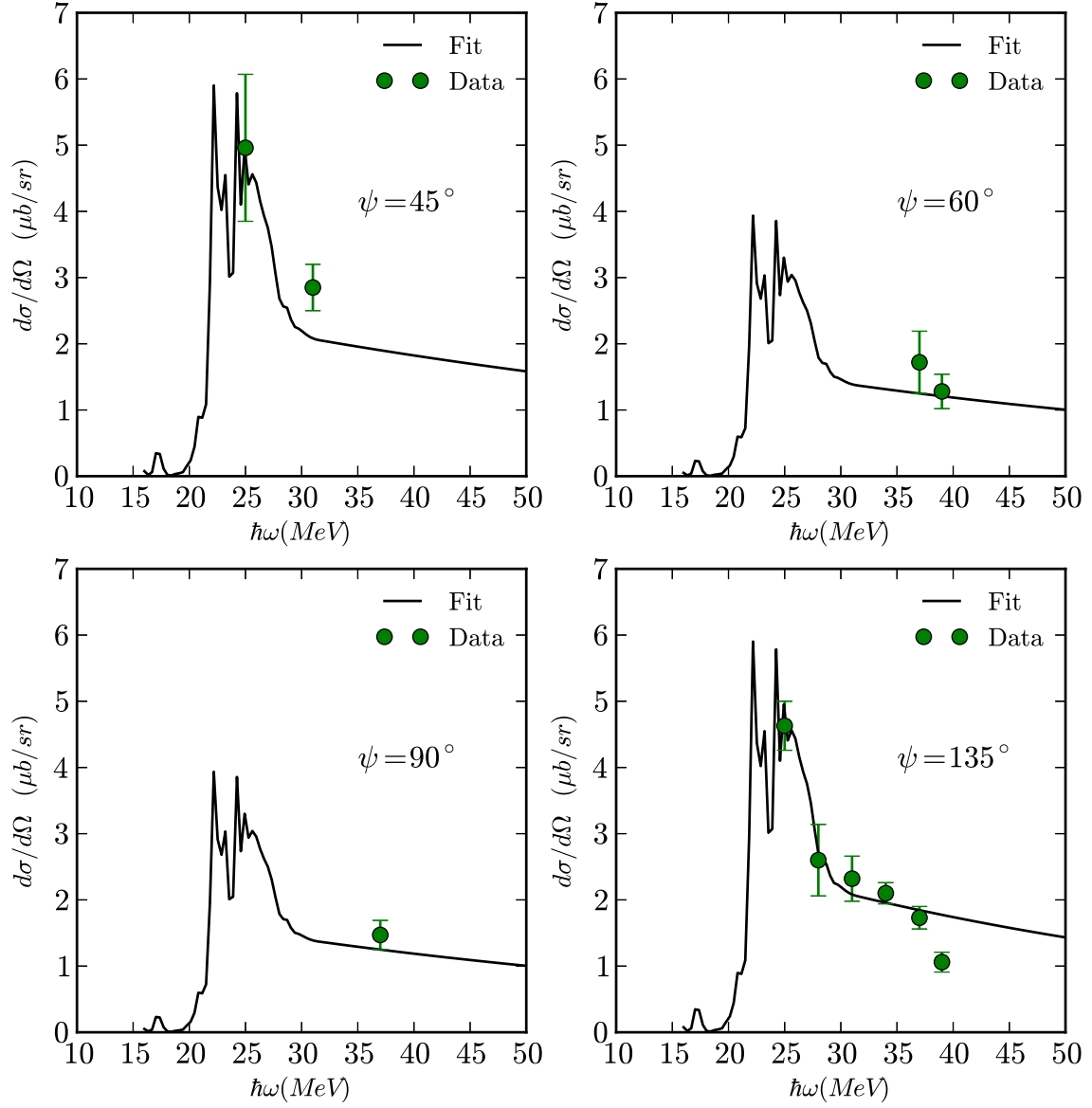}
\caption{\textsl{Comparison of our calculation, dark line, with data for $^{16}O$ from \cite{Hayward83} for four different scattering angles.}}
\label{fig:Oxygen}
\end{center}
\end{figure*}

The outcome of our model is included in Figures \ref{fig:oldPbfit}--\ref{fig:Oxygen}.
As is apparent from Fig. \ref{fig:oldPbfit}, for the case of lead the quality of the above fit is comparable to that appearing in \cite{Sore10} and cited in Eq. (\ref{eq:Pbfit}) except for a considerably higher quality at 60 degrees which we assume to be a general small-angle feature. 
For the case of carbon, Figures \ref{fig:compareCarbon} and \ref{fig:compareCarbonHighE} reveal that our simple procedure yields quite reasonable cross sections with deviations from experimental data typically within 10 \% for all angles at energies beyond the giant dipole resonance. 
Note that with a dividing energy of $\hbar\tilde{\omega}_2 = 32.8$ MeV, Eq. (\ref{eq:valuedividingE}), our curves in Fig. \ref{fig:compareCarbon} are produced solely by use of the photonuclear absorption cross section and the optical theorem, the dispersion relation, and a dipole radiation pattern, Eqs. (\ref{eq:Im}--\ref{eq:gdipole}), whereas the damping introduced in this subsection enters critically in Fig. \ref{fig:compareCarbonHighE}.
Our procedure should not be used beyond the pion-production threshold and in the determination of $\epsilon$ and $\delta$ we have focused on energies up to about 80--100 MeV since higher values contribute little to the bremsstrahlung. 
Although scarce, some data on elastic scattering of photons on oxygen exist. 
We show these data along with our calculations in Figure \ref{fig:Oxygen}.
The agreement is good even though our approach has only been tuned using C and Pb. 


With the above procedure, it is now possible to calculate the bremsstrahlung spectrum from any bare relativistic ion for which the photon absorption cross section is known. 
In this paper, we use values for the latter available from \cite{ENDF}.

\section{Bremsstrahlung for any bare relativistic ion - results}
By application of the procedure described in Sec. \ref{subsec:newapproach} the calculation of the bremsstrahlung spectrum pertaining to the penetration of a thin foil can be carried out for any incoming relativistic bare ion for which the total photonuclear absorption cross section is known. 

As a prelude to the general case, we first perform calculations for $^{208}_{82}$Pb in order to check against our previous results. Ref. \cite{Sore10} presents results for $^{208}_{82}$Pb ions penetrating a lead target at $\gamma = 170$. These are repeated in Fig. \ref{fig:BSPb170} (solid curve). By application of  the procedure described in Sec. \ref{subsec:newapproach}, the elastic photonuclear cross section is determined for all angles according to our new and general scheme supposed to be valid for any ion. Our new estimate of the bremsstrahlung spectrum is obtained by substituting this cross section into Eq. (\ref{eq:doubleDiff}) and performing the steps described subsequently in Sec. \ref{sec:BS}. The result 
is included in Fig. \ref{fig:BSPb170} (dotted curve). 
The new procedure closely reproduces the result of Ref. \cite{Sore10} which is based on the fit (\ref{eq:Pbfit}).

\begin{figure}[hbt]
\begin{center}
\includegraphics[width=0.5\textwidth]{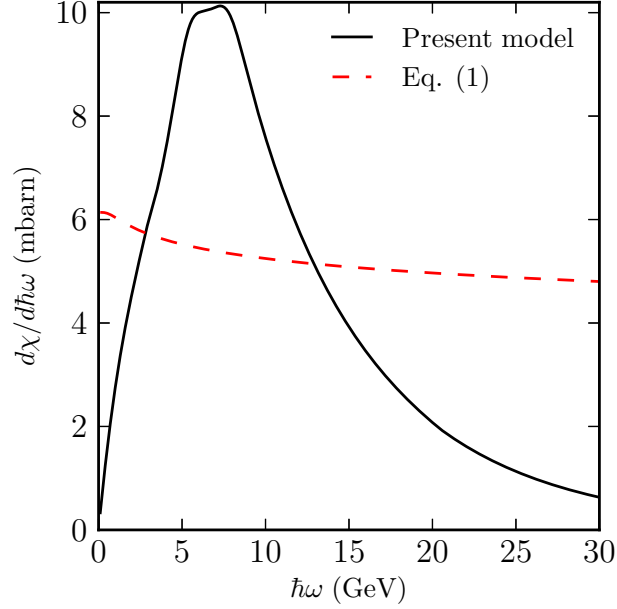}
\caption{\textsl{Bremsstrahlung from a $^{40}$Ar projectile with $\gamma$ = 170 incident on a lead target. The solid dark line shows the present calculations and the dashed red line shows the reference cross section (\ref{eq:Reference}).}
\label{fig:gamma170}}
\end{center}
\end{figure}

As an illustration of the general case, we take $^{40}_{18}$Ar ions traversing a radiation target made of lead. Except for the effect of atomic screening of the target nucleus, cross sections for other targets can be found by a simple scaling, see \cite{Sore10}. 
The bremsstrahlung cross section is shown in Fig. \ref{fig:gamma170} along with the reference cross section, Eq. (\ref{eq:Reference}). 
As was the case for lead, the bremsstrahlung spectrum displays a resonant structure due to the giant dipole resonance. 
The giant dipole resonance manifests itself in the bremsstrahlung spectrum as a peak located at an energy that is shifted upwards by a factor of $2 \gamma$. 
For argon, the peak is rather broad compared to the bremsstrahlung spectrum for lead. 
This is caused by the shape of the argon nucleus which is less spherically symmetric than lead, and therefore the photonuclear absorption cross section for argon has $2$ overlapping giant dipole resonance peaks. 
The peak in the bremsstrahlung spectrum therefore consists of 2 closely located peaks. 
As for lead, the bremsstrahlung emission decreases gradually towards high energies, with an exponential-like shape reflecting the exponential decrease in elastic photon scattering cross section. 
The reference cross section, Eq. (\ref{eq:Reference}), remains almost constant all the way up to the projectile energy. 

\begin{figure}[hbt]
\begin{center}
\includegraphics[width=0.5\textwidth]{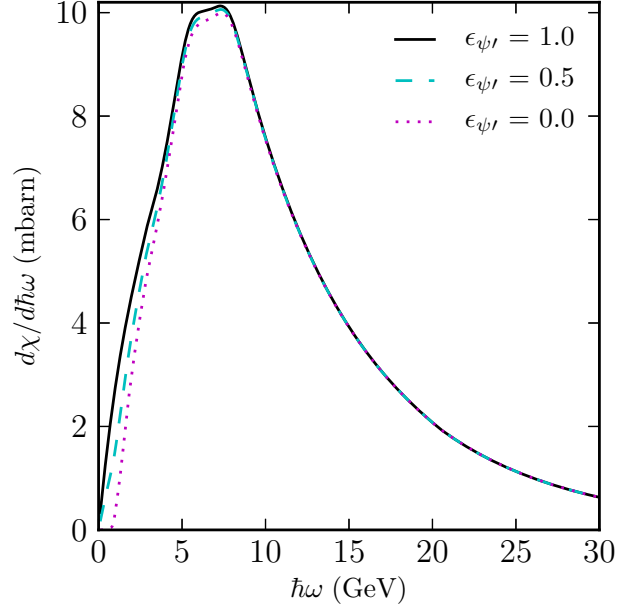}
\caption{\textsl{Bremsstrahlung from a $^{40}$Ar projectile with $\gamma$ = 170 incident on a lead target. The solid line shows the full spectrum as in Figure \ref{fig:gamma170}, the dashed line shows the spectrum when the cross section for scattering angles smaller than $\pi /4$ has been reduced by a factor of $0.5$, and the dotted line shows the spectrum when small-angle scattering has been completely removed.}
\label{fig:gamma170cutBelow}}
\end{center}
\end{figure}

It is evident that the bremsstrahlung spectrum is dominated by the elastic scattering of virtual photons with energies near the giant dipole resonance. 
Also, the spectrum is mainly due to virtual photons which scatter in the forward direction, $\theta' \approx 0$ or $\psi' \approx \pi$, and hence gain most energy by the transformation to the laboratory frame. Figure \ref{fig:gamma170cutBelow} illustrates this point.
On the figure, we show the full spectrum together with the result of calculations where the contribution from small-angle scattering, specifically angles $\psi' $ smaller than $\pi /4$, has been artificially reduced by a factor of $\epsilon_{\psi'}=0.5$, or even removed completely, $\epsilon_{\psi'}=0$.  
Even when small-angle scattering is completely removed, the shape of the bremsstrahlung spectrum is changed only slightly and only at energies below the peak. The reduction does not change the location of the peak. The relatively weak sensitivity to small-angle scattering implies a relatively large tolerance of inaccuracies in the photo nuclear cross section for such angles. It may be noted that, in any case, a substantial reduction of the spectrum on the low-energy side of the peak is obtained by forward collimation to angles $\theta$ assuming values of only a fraction of $1/\gamma$.

\begin{figure}[hbt]
\begin{center}
\includegraphics[width=0.5\textwidth]{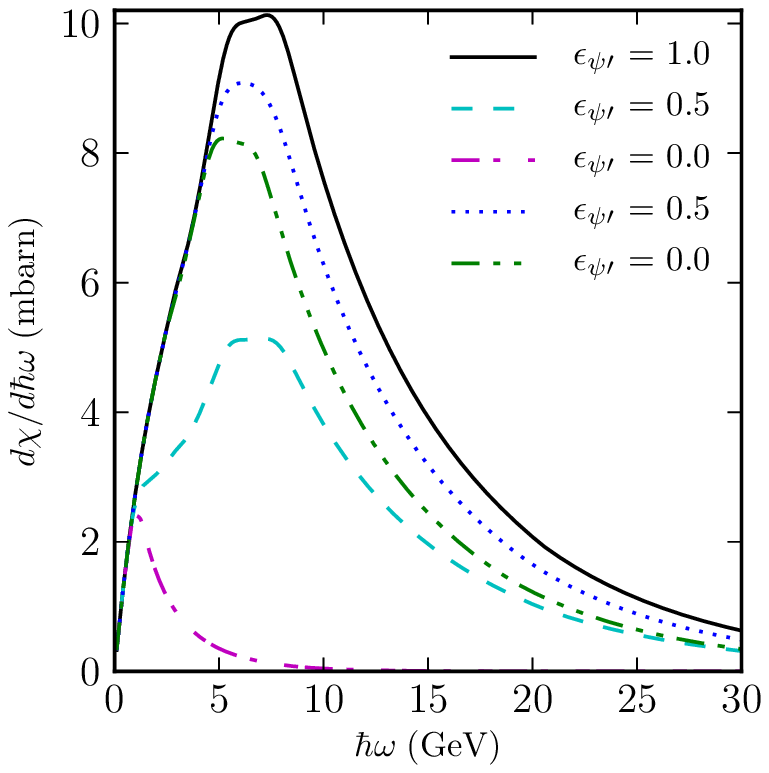}
\caption{\textsl{Bremsstrahlung from a $^{40}$Ar projectile with $\gamma$ = 170 penetrating a lead target. The solid line shows the full spectrum as in Figure \ref{fig:gamma170} together with the result of calculations where the large-angle elastic scattering has been removed or dampened. In the legend, the $3$ upper entries are for $\psi ' > \pi / 4$ and the $2$ lower entries are for $\psi ' > 3\pi /4$.}
\label{fig:gamma170cutAbove}}
\end{center}
\end{figure}

Figure \ref{fig:gamma170cutAbove} shows the role played by large-angle scattering in the composition of the bremsstrahlung spectrum. Results are displayed for the case when the contribution from scattering angles $\psi' $ larger than $\pi /4$ and $3\pi /4$ has been artificially halved or removed completely by applying an angular dependent damping factor $\epsilon_{\psi '}$ to the cross section.
By reducing the large-angle contribution, the bremsstrahlung spectrum is not only altered at energies above the peak, but the peak height and location are also changed dramatically. 
This shows the importance of modelling the large-angle scattering carefully.

\section{Summary and Conclusions}
We have presented calculations of the bremsstrahlung emission originating from ultrarelativistic bare heavy ions penetrating an amorphous material. 
Using a Weizs\"{a}cker-Williams approach, we have incorporated existing knowledge on the photonuclear interactions to obtain a more realistic bremsstrahlung spectrum than what results for a pointlike particle of the same mass and charge as the actual ion.
The latter here serves as a reference. 
Overall, the spectra are dramatically different for the extended and the pointlike projectile. 
But the height of the peak for the composite nucleus and the almost constant value of the reference cross section (\ref{eq:Reference}) scale roughly similarly with atomic number: at $\gamma$ = $170$ the peak-value overshoots the reference cross section by a factor of about 6 for lead and a factor of about 2 for argon.
For energies below and above the peak, i.e. below and above $5-10$ GeV for $\gamma$ = $170$, our calculations show how the traditional bremsstrahlung calculations significantly overestimate the cross section.
This is especially true on the high-energy side of the peak, where our calculations show an exponential decrease in cross sections. 

The giant dipole resonance directly translates to a resonance structure in the bremsstrahlung spectrum which is located at $2\gamma$ times the giant dipole resonance energy. The giant dipole resonance is typically located at energies around $10-20$ MeV. 
For relativistic argon ions with $\gamma = 170$, the corresponding peak in the bremsstrahlung spectrum appears at about $5$ GeV.  

Our calculations can be extended to all ions for which the total absorption cross section is known and the calculations can be performed for any incoming beam energy and target material. 
The bremsstrahlung emission from heavy bare argon and lead ions may be measured at the SPS during 2015-2017. 

The shape of the bremsstrahlung peak reflects the shape of the giant dipole resonance. Hence the structure of any nucleus may be exposed via the detection of bremsstrahlung photons; much like the obvious statement that any nucleus can be studied using a photon beam. 
But there are some advantages to the highly Lorentz boosted environment discussed here. 
One may for example study unstable nuclei whose lifetimes are too short to aim a photon beam at them. 
This would make it possible to study radioactive elements and exotic nuclei. 
Another benefit is that the bremsstrahlung photons are emitted in the forward direction and can easily be measured. 
In a traditional scattering experiment where the ion is at rest, the regions corresponding to photons scattered at $0^\circ$ and $180^\circ$ are normally not possible to instrument. 


\end{document}